\documentclass[twocolumn,pra]{revtex4}
\usepackage{amssymb}
\usepackage{amsmath}
\usepackage{txfonts}
\usepackage{amsfonts}
\usepackage{graphicx}

\begin{document}

\title{Antibunched photon-pair source based on photon blockade in a
nondegenerate optical parametric oscillator}
\author{Yi Ren, Shouhui Duan, Wenzhi Xie,
Yongkang Shao, and Zhenglu Duan$^{\ast }$}
\affiliation{College of Physics, Communication and Electrons, Jiangxi Normal University,
Nanchang, 330022, China }

\begin{abstract}
Nonclassical light sources, such as correlated photon-pairs, play an
important role in quantum optics and quantum information processing systems.
This study proposes a process to generate antibunched photon-pairs in a
nondegenerate optical parametric oscillator. It is found that when the
parameters of the system satisfy certain conditions, the generated photons
in subharmonic modes exhibit a strong antibunching behavior and are strongly
correlated with one another. In particular, the average photon-pair number
is resonantly enhanced. It is also observed that the conventional photon
blockade contributes to this phenomenon. In addition, it is interesting to
note that fundamental mode photons can blockade the subharmonic mode
photons. We refer to this phenomenon as a heterogeneous photon blockade.
\end{abstract}

\maketitle
\footnotetext[1]{$^\ast$Corresponding author, duanzhenglu@jxnu.edu.cn}

The efficient generation of correlated photon-pairs over long time periods
is of interest in the field of quantum optics, which is widely used in
physics, chemistry, and biology in applications such as two-photon
excitation microscopes \cite{TPM}, multiphoton ionization\cite{MPA}, the
testing of fundamental quantum mechanics\cite{QMtest}, the implementation of
quantum communications and computing\cite{QCM,Qcomputer}, and quantum
imaging. Therefore, generating high-performance correlated photon-pairs has
attracted the attention of researchers\cite{16}, and it is especially
desired in quantum optics labs. Currently, there are two major methods used
to generate correlated photon pairs. One is based on nonlinear parametric
processes, such as the spontaneous parametric down-conversion process (SPDC)%
\cite{SPDC}/optical parametric oscillator (OPO)\cite{OPO} in a nonlinear
crystal or spontaneous four-wave mixing (SFWM) in a fiber\cite{fiber} and
atomic ensemble\cite{atom}. The other involves implementing multilevel
atomic cascades or biexcitation decays of quantum dots. Considering a
nonlinear parametric process, the generated photon-pairs have some
drawbacks. These include the statistical generation of pairs and the
emission and broad spectral properties of multiphotons, which limit the
applications of pairs from SPDC or SFWM in quantum networks\cite{Qnetwork}.
To overcome these drawbacks, a method involving spectral filtering and a low
pump power supplied to the source of photon-pairs is used, but this results
in a very low generation rate of pairs per pump pulse. Therefore, new
schemes to generate high-performance correlated photon-pairs are required.

The photon blockade (PB) effect can convert a classic light with a
Poissonian or sup-Poissonian distribution into a nonclassic light with a
sub-Poissonian distribution, and it has been used to generate high purity
single-photon sources\cite{PB}. The PB effect can be divided into
conventional photon blockades (CPBs)\ and unconventional photon blockades
(UPBs)\cite{UPB,UPB1}. The former is based on the anharmonicity of the
energy level spacing of the system. When a certain energy level is excited
by external drive resonance, the detuning caused by the anharmonicity of the
energy level will inhibit the excitation of other energy levels, so as to
improve the purity of the light field with a specific photon number. The
latter uses destructive quantum interference to suppress specific energy
levels and improve the purity of the photon source.

The strong nonlinearity of the system can arise from a single two-level atom
that is strongly coupled to an optical cavity mode\cite{TLS_PB,TLS2} or from
a three-level emitter coupled to a cavity operating near the
electromagnetically induced transparency window\cite{EIT}. CPB was first
observed experimentally in a system with a trapped atom in an optical cavity%
\cite{CPBEx}. In addition to atom--cavity systems, other quantum systems,
such as quantum-dot--cavity systems\cite{QDC}, superconducting qubit systems%
\cite{SQS}, and optomechanical systems\cite{OMS,OMS1}, have been used to
explore CPB and serve as single photon sources. Motivated by the above
discussion, we explore the generation of antibunching photon-pairs based on
the conventional photon blockade mechanism in this study.

The model we consider in this study is a three-mode optical cavity filled
with $\chi ^{(2)}$ nonlinear media that mediates the conversion of a photon
in the fundamental mode $a$ at frequency $\omega _{a}$ to two photons in the
subharmonic modes $b$ at frequency $\omega _{b}$ and $c$ at frequency $%
\omega _{c}$, and vice-versa. The fundamental mode $a$ is pumped by an
external driving light at frequency $\omega _{l}$ with amplitude $E$, which
forms a nondegenerate optical parametric oscillator \cite%
{SONSM,SONSM1,SONSM2,SONSM3}. The model is schematically shown in Fig. 1(a).
In the rotating frame with respect to the driving light frequency, the
Hamiltonian can be given by $\left( \hbar =1\right) $

\begin{eqnarray}
\hat{H} &=&\Delta _{a}\hat{a}^{\dag }\hat{a}+\Delta _{b}\hat{b}^{\dag }\hat{b%
}+\Delta _{c}\hat{c}^{\dag }\hat{c}  \label{Heff} \\
&&+g\left( \hat{a}\hat{b}^{\dag }\hat{c}^{\dag }+\hat{a}^{\dag }\hat{b}\hat{c%
}\right) +E\left( \hat{a}^{\dag }+\hat{a}\right) ,  \notag
\end{eqnarray}%
where $\hat{a}(\hat{a}^{\dag })$, $\hat{b}(\hat{b}^{\dag })$, and $\hat{c}(%
\hat{c}^{\dag })$ are the annihilation (creation) operators of the cavity
modes. $\Delta _{a}=\omega _{a}-\omega _{l}$ and $\Delta _{b,c}=\omega
_{b,c}-\omega _{l}/2$ represent the detuning between the cavity modes and
the driving field. Next, we mainly focus on the quantum statistics of
strongly correlated photon-pairs in modes $b$ and $c$.
\begin{figure}[tbp]
\includegraphics[width=3.2in]{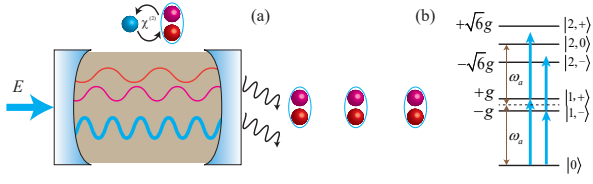}
\caption{{\protect\footnotesize (Color online) (a) Nondegenerate SPDC
process induced by a second-order nonlinear medium demonstrated in a
three-mode optical cavity, which is driven by a weak coherent light. (b)
Dressed state energy level transition diagram of the system. When the
external driving frequency resonates with one of the dressed states, it will
detune with the dressed states of the other energy levels, thereby
inhibiting the system from entering other states and forming CPB.}}
\end{figure}

To quantitatively study the quantum statistics of the generated
photon-pairs, we use the state vector method to analyze the quantum
statistics of the photon-pairs. We phenomenologically introduce the cavity
decay in the non-Hermitian Hamiltonian:%
\begin{equation}
\hat{H}_{\text{non}}=\hat{H}-\frac{i\kappa _{a}}{2}\hat{a}^{\dag }\hat{a}-%
\frac{i\kappa _{b}}{2}\hat{b}^{\dag }\hat{b}-\frac{i\kappa _{c}}{2}\hat{c}%
^{\dag }\hat{c},
\end{equation}%
where $\kappa _{a}$($\kappa _{b}$, $\kappa _{c}$) is the decay rate of modes
$a$($b$, $c$). In the weak driving limit, i.e., $E\ll \{\kappa _{a,b,c}\}$,
the wave function of the system is approximately expressed as:%
\begin{eqnarray}
\left\vert \psi \right\rangle &=&C_{000}\left\vert 000\right\rangle
+C_{100}\left\vert 100\right\rangle +C_{011}\left\vert 011\right\rangle
\notag \\
&&+C_{200}\left\vert 200\right\rangle +C_{111}\left\vert 111\right\rangle
+C_{022}\left\vert 022\right\rangle .
\end{eqnarray}%
Here, $\left\vert mnl\right\rangle $ is the Fock-state basis of the system
where $m$ denotes the photon number in the fundamental mode $a$, and $n$ and
$l$ denote the photon numbers in the subharmonic modes $b$ and $c$,
respectively. From the Schrodinger equation, we can obtain the evolution
equations of the amplitude of the wave function. The analytical results of
the probability amplitude are obtained by using the steady state assumption.

\begin{figure}[tbp]
\includegraphics[width=3.4in]{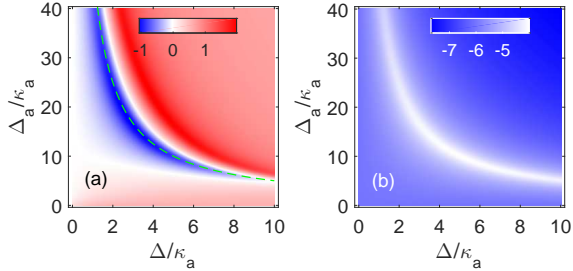}
\caption{{\ (Color online) (a) Auto-correlation function of a photon-pair as
a function of the detunings $\Delta _{a}$ and $\Delta _{b}$, which is based
on a logarithm function of 10. (b) Average photon-pair number as a function
of the detunings $\Delta _{a}$ and $\Delta _{b}$. The parameters are set as $%
g=10\protect\kappa _{a}$, $\protect\kappa=\protect\kappa _{a}/2$, and $E=0.01%
\protect\kappa _{a}$.}}
\end{figure}

In the NOPO, one photon in the fundamental mode $a$ will simultaneously be
converted into two photons in the second harmonic modes $b$ and $c$, which
can be considered as a two-photon state. To describe the two-photon state,
we define $\hat{D}=\hat{b}\hat{c}$ $(\hat{D}^{\dag }=\hat{b}^{\dag }\hat{c}%
^{\dag })$ as the annihilation (creation) operator for the two-photon state
or photon--pair state. $n_{D}\equiv \left\langle \hat{D}^{\dag }\hat{D}%
\right\rangle \approx \left\vert C_{0,1,1}\right\vert ^{2}$ is the average
intensity of the photon-pair, which can be used to characterize the
brightness of the photon pair. $g_{D}^{\left( 2\right) }\left( 0\right).
\equiv \left\langle \hat{D}^{\dag 2}\hat{D}^{2}\right\rangle /\left\langle
\hat{D}^{\dag }\hat{D}\right\rangle ^{2}\approx \left\vert
C_{0,2,2}\right\vert ^{2}/\left\vert C_{0,1,1}\right\vert ^{4}$ represents
the auto-correlation function of the photon pair, which can be used to
determine the quantum statistics of the photon-pair, such as the
antibunching feature. After completing the straightforward calculation, we
obtain the analytical expression for the auto-correlation function for the
photon-pair as follows:%
\begin{equation}
g_{D}^{\left( 2\right) }\left( 0\right) \approx 4\left\vert 1-\frac{%
g^{2}\Delta _{a}^{\prime }}{\left( \Delta _{a}^{\prime }+\Delta
_{b}^{\prime. }+\Delta _{c}^{\prime }\right) \left( \left( \Delta
_{b}^{\prime }+\Delta, _{c}^{\prime }\right) \Delta _{a}^{\prime
}-g^{2}\right) }\right\vert ^{-2},  \label{g2D}
\end{equation}%
and the average photon-pair number may be defined as:%
\begin{equation}
n_{D}\approx \frac{g^{2}E^{2}}{\left\vert \left( \Delta _{b}^{\prime
}+\Delta _{c}^{\prime }\right) \Delta _{a}^{\prime }-g^{2}\right\vert ^{2}},
\label{nD}
\end{equation}%
Here, $\Delta _{a}^{\prime }=\Delta _{a}-i\kappa _{a}/2$ and $\Delta
_{b,c}^{\prime }=\Delta _{b,c}-i\kappa _{b,c}/2$. From Eqs. (\ref{g2D}) and (%
\ref{nD}) it can be observed that the self-correlation function and average
number of photon pairs are dependent on the sum of the detunings $\Delta
_{b} $ and $\Delta _{c}$ and are not dependent on themselves. This feature
relaxes the condition required to tune the quantum statistics of
photon-pairs. For simplicity and without the loss of physics, we let the
cavity detunings and decay rates for modes $b$ and $c$ be equivalent such
that $\Delta _{b}=\Delta _{c}=\Delta $ and $\kappa _{b}=\kappa _{c}=\kappa $.

It is known that a vanishing auto-correlation function corresponds to an
ideal antibunching of a photon pair. That is, a photon pair with antibunched
features will prefer not to be in close proximity to other photon pairs.
From Eq. (\ref{g2D}), if $\left\vert \left( \Delta _{b}^{\prime }+\Delta
_{c}^{\prime }\right) \Delta _{a}^{\prime }-g^{2}\right\vert =0$ and the
second order correlation function is zero, these conditions correspond to an
ideal antibunching of a photon pair. However, this condition does not hold
in real-world physical situations. At the strong coupling limit $g\gg
\{\kappa ,\kappa _{a}\}$, the condition reduces to:

\begin{equation}
2\Delta \Delta _{a}=g^{2},  \label{opt}
\end{equation}%
which can minimize the auto-correlation function and results strong
photon-pairs that exhibit antibunching with appropriate parameters. When the
reduced optimal condition (Eq. (\ref{opt})) holds, the average photon-pair
numbers are resonantly enhanced. Eq. (\ref{opt}) appears to represent the
condition for maximizing the average photon-pair number. This phenomenon is
similar with that observed for a conventional single-photon blockade.
Therefore, we refer to this phenomenon as a photon-pair blockade.
Additionally, we find that the reduced optimal conditions (Eq. (\ref{opt}))
are similar to those in the case of three wave mixing \cite{SPDC0}.

To explain the photon-pair blockade discussed above, we analyze the
transition of the dressed states. Similar to the procedure used in ref. \cite%
{SPDC0}, considering the strong-coupling limit, we determined the total
number of photonic excitations of the system, $n$, is a conserved quantity
under Hamiltonian action without considering the external driving force. In
this case, the energy eigenstates of the Hamiltonian can be classified into $%
n$-manifolds with dressed states. The $1$-manifold of the energy level
consists of a doublet $\left\vert 1,\pm \right\rangle =\left( \left\vert
011\right\rangle \pm \left\vert 100\right\rangle \right) /\sqrt{2}$ with
eigenenergy $E_{1,\pm }=[\omega _{a}+\omega _{b}+\omega _{c}\pm \sqrt{%
4g^{2}+\left( \omega _{a}-\omega _{b}-\omega _{c}\right) ^{2}}]/2$. If the
transitions between the ground state $\left\vert 000\right\rangle $ and
doublets $\left\vert 1,\pm \right\rangle $ are resonantly excited (as in $%
\omega _{l}=E_{1,\pm }$), then $2\Delta \Delta _{a}=g^{2}$. This is the
reduced optimal condition (Eq. (\ref{opt}). For the case of $\omega
_{b}+\omega _{c}=\omega _{a}$, the $2$-manifold of the energy level consists
of a triplet $\left\vert 2,\pm \right\rangle =(\left\vert 200\right\rangle +%
\sqrt{3}\left\vert 022\right\rangle \pm \sqrt{2}\left\vert 111\right\rangle
)/\sqrt{6}$ with eigenenergy $2\omega _{a}\pm \sqrt{6}g$ and $\left\vert
2,0\right\rangle =(\sqrt{2}\left\vert 200\right\rangle -\left\vert
022\right\rangle )/\sqrt{3}$ with eigenenergy $E_{2,0}=2\omega
_{a}\allowbreak $. In this case, the transitions from $1$-manifold to $2$%
-manifold ($\left\vert 1,\pm \right\rangle \leftrightarrow \left\{
\left\vert 2,\pm \right\rangle ,\left\vert 2,0\right\rangle \right\} $) is
suppressed due to far off-resonance with large detunings in the
strong-coupling limit. Therefore, there is at most one photon in the
fundamental mode $a$ or a pair of photons in the subharmonic modes $b$ and $%
c $ in the cavity simultaneously.

To further quantitatively demonstrate the quantum statistics of the
photon-pairs described above, we utilize a numerical simulation based on the
quantum master equation:%
\begin{equation}
\frac{\partial \hat{\rho}}{\partial t}=-i[\hat{H}_{eff},\hat{\rho}]+\frac{%
\kappa _{a}}{2}L\left[ \hat{a}\right] \hat{\rho}+\frac{\kappa }{2}L[\hat{b}]%
\hat{\rho}+\frac{\kappa }{2}L\left[ \hat{c}\right] \hat{\rho},
\end{equation}%
where $L[\hat{A}]\hat{\rho}=2\hat{A}\hat{\rho}\hat{A}^{\dag }-\hat{A}^{\dag }%
\hat{A}\hat{\rho}-\hat{\rho}\hat{A}^{\dag }\hat{A}$ is the Lindblad term
accounting for losses to the environment. The zero-delay auto-correlation
function and cross-correlation function can be expressed as $g_{x}^{\left(
2\right) }\left( 0\right) =Tr(x^{\dag 2}x^{2}\hat{\rho}_{s})/Tr(x^{\dag }x%
\hat{\rho}_{s})^{2}$ ($x=\hat{a},\hat{b},\hat{c},\hat{D}$) and $%
g_{xy}^{\left( 2\right) }\left( 0\right) =Tr(x^{\dag }xy^{\dag }y\hat{\rho}%
_{s})/[Tr(x^{\dag }x\hat{\rho}_{s})Tr(y^{\dag }y\hat{\rho}_{s})]$ ($x=\hat{a}%
,\hat{b},\hat{c},\hat{D}$), respectively, when the system tends towards the
steady state $\hat{\rho}_{s}\equiv \hat{\rho}(t\rightarrow \infty )$. The
average photon-pair number is $n_{D}=Tr(\hat{D}^{\dag }\hat{D}\hat{\rho}%
_{s}) $. In the numerical simulation, we truncated the photon number space
to 40, which safely guarantees the convergence of the simulation.

In Fig. 2, we show the auto-correlation function of photon pairs and the
average photon-pair number with changing cavity detunings $\Delta _{a}$ and $%
\Delta $. The autocorrelation function $g_{D}^{\left( 2\right) }\left(
0\right) $ shows a dark blue region less than $1$, which indicates that the
photon-pairs in the cavity follow the sub-Poisson distribution. The green
dashed line is plotted using the reduced optimal condition (Eq. (\ref{opt}),
and it indicates the local optimal antibunching of photon-pairs. It is
evident that the reduced optimal condition is very consistent with the
numerical results. Meanwhile, we observe the average photon-pair number as
it is resonantly enhanced along the reduced optimal curve (green dashed
line) in Fig. 2(b). Therefore, the NOPO has the potential to serve as a
high-brightness and high-purity photon-pair source.

We further study the correlation of the photon-pairs when the parameters of
the system are those of the reduced optimal condition. Fig. 3(a) shows $%
g_{D}^{\left( 2\right) }\left( 0\right) $ and $n_{D}$, which change with the
cavity detuning $\Delta _{a}$ for $\Delta =g^{2}/2\Delta _{a}$. It can be
observed that for small $\Delta _{a}$ values, the auto-correlation is
greater than one, which corresponds to photon-pair bunching, whereas for $%
\Delta _{a}\in (g/2,20g)$, the auto-correlation is less than one and the
smallest $g_{D\min }^{\left( 2\right) }\left( 0\right) \sim 0.02$ value
occurs at $\Delta _{a}\approx \sqrt{3}g$, implying strong photon-pair
antibunching. It can also be observed that the average photon-pair number
first increases and then decreases as the detuning $\Delta _{a}$ is
increased. The largest value of the average photon-pair number occurs at $%
\Delta _{a}\approx g$.

To analyze the aforementioned phenomenon, we rewrite the self-correlation
function of photon-pairs using the reduced optimal condition (Eq.(\ref{opt}%
)):
\begin{equation}
g_{D}^{\left( 2\right) }\left( 0\right) \approx \frac{4\left( \Delta
_{a}^{2}+g^{2}\right) ^{2}\left( g^{2}\kappa _{a}+\Delta _{a}^{2}\kappa
_{D}\right) ^{2}}{\kappa _{a}^{2}g^{8}+4g^{4}\Delta _{a}^{6}}.  \label{apG2D}
\end{equation}%
which has a minimum value at $\Delta _{a,\text{optimal}}\approx g\sqrt{%
[\kappa _{a}/\kappa +2+\sqrt{\kappa _{a}^{2}/\kappa ^{2}+28\kappa
_{a}/\kappa +4}]/4}$. Because we set $\kappa _{a}=2\kappa $ in the numerical
simulation, then $\Delta _{a,\text{optimal}}\approx \sqrt{3}g$. Considering
the average photon-pair number based on the analytical expression in Eq. (%
\ref{nD}), we find that the optimal detuning for the maximum $n_{D}$ is $%
\Delta _{a}=g $. These analytical results agree well with the numerical
results discussed above.

To further understand the antibunching of the photon-pair generated in NOPO,
we also plot the self-correlation functions of the fundamental mode $a$ and
subharmonic modes $b$ and $c$ in Fig. 3. It can be seen that the photons in
the fundamental and subharmonic modes, respectively, exhibit stronger
antibunching features around $\Delta _{a}\sim g$ using the reduced optimal
condition (Eq.(\ref{opt}). Specifically, we find that $g_{b,c}^{\left(
2\right) }\left( 0\right) \ll g_{D}^{\left( 2\right) }\left( 0\right) $.
This can be understood as indicating that one photon-pair may become a
single-photon when another photon leaks out of the cavity.

Fig. 3 also shows the cross-correlation between the subharmonic modes $b$
and $c$. It should be noted that $g_{bc}^{\left( 2\right) }\left( 0\right)
\gg 1$, indicating a very strong bunching between photons $b$ and $c$. This
phenomenon is evident because the fundamental mode photon $a$ is constantly
being simultaneously converted into pair-photons in modes $b$ and $c$.
Considering the cross-correlation functions between the photons in the
fundamental and subharmonic modes, a different scenario is observed. It can
be seen that $g_{ab,ac}^{\left( 2\right) }\left( 0\right) \ll 1$, indicating
that a strong anticorrelation between the fundamental and subharmonic modes
exists. Therefore, if there is a fundamental mode photon $a$ in the cavity,
there are no subharmonic mode photons $b$ or $c$, and vice-versa. This
occurs because other photons in the fundamental mode cannot enter into the
cavity when a coherent conversion between photons in the fundamental and
subharmonic modes occurs in the cavity.

\begin{figure}[tbp]
\includegraphics[width=3.2in]{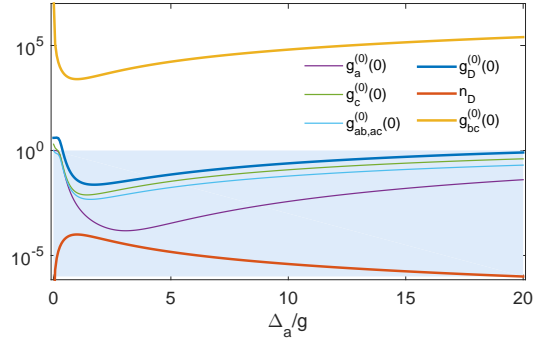}
\caption{{(Color online) Properties of correlated photon-pairs as a function
of $\Delta _{a}$. The fixed coupling strength is $g=10\protect\kappa _{a}$,
and the other parameters are set as $\protect\kappa=\protect\kappa _{a}/2$, $%
\Delta=g^{2}/(2\Delta_{a})$, and $E=0.01\protect\kappa _{a}$.}}
\end{figure}

We also examine the dependence of the auto-/cross-correlation functions on
the nonlinear interaction strength $g$, as shown in Fig. 4(a), as well as
the decay rate of the subharmonic modes $\kappa $, as shown in Fig. 4(b). It
is observed that the cross-correlation function $g_{bc}^{\left( 2\right)
}\left( 0\right) $ is almost unchanged while the auto-correlation function
of the photon-pairs decreases with increasing coupling strength. This
indicates that stronger coupling can help improve the purity of the
photon-pairs but not the correlation between photons in the subharmonic
modes. This occurs because a stronger coupling strength indicates a larger
detuning between the dressed state, $\left\vert 2,\pm \right\rangle $, and
the driving frequency, which results in a stronger suppression of the
population of the dressed states $\left\vert 2,\pm \right\rangle $ and $%
\left\vert 2,0\right\rangle $. Considering the dependences of the
correlation function and the average photon-pair number on the decay rate of
the subharmonic modes, a different scenario is observed. All of the
auto-/cross-correlation functions are observed to increase and the average
photon-pair number decreases with an increasing the decay rate. A small
mode-decay rate results in the generation of high-purity and high-intensity
photon-pair sources. This occurs because a large decay rate leads to a
larger probability of transition into a double photon-pair state $\left\vert
2,\pm \right\rangle $ and $\left\vert 2,0\right\rangle $ from a single
photon-pair state $\left\vert 1,\pm \right\rangle $.

\begin{figure}[tbp]
\includegraphics[width=3.2in]{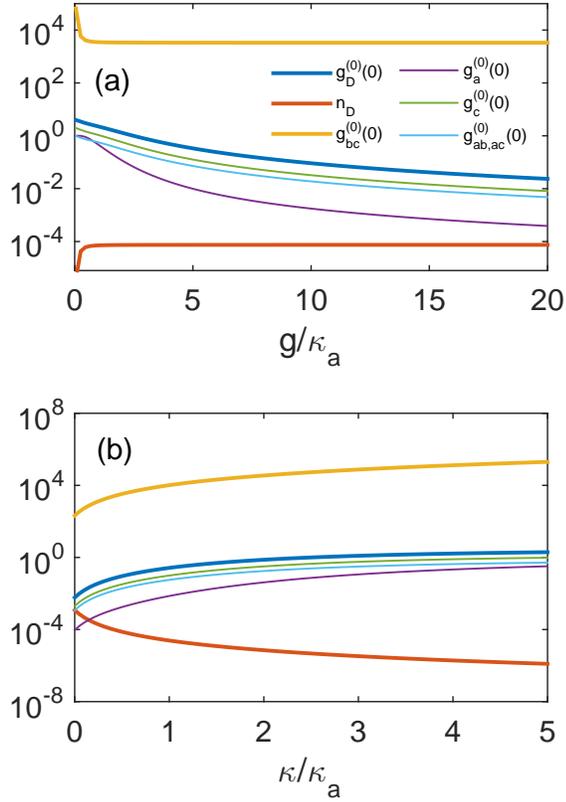}
\caption{{(Color online) Property changes of correlated photon-pairs with
the (a) coupling strength $g$ with $\protect\kappa=\protect\kappa _{a}/2$ or
(b) decay of subharmonic mode $\protect\kappa$ with $g=10\protect\kappa_a$.
The fixed coupling strength is $\Delta _{a}=g/\protect\sqrt{3}$. The other
parameters are set as $\Delta=g^{2}/(2\Delta _{a})$ and $E=0.01\protect%
\kappa _{a}$.}}
\end{figure}

In conclusion, in this work we proposed a schematic model for the
preparation of an antibunched photon-pair source in an NOPO based on the
conventional photon blockade effect. The quantum statistics of the
photon-pairs at steady state are analyzed under a continuous weak driving
force. It is concluded that the system can realize the preparation of
photon-pairs with a strong cross-correlation between the photons in the
pairs and a very small auto-correlation ($g_{D}^{\left( 2\right) }\left(
0\right) \ll 1$) with a strong $\chi ^{(2)}$ nonlinear coupling strength.
These qualities are indicative of a high-purity and high-intensity
photon-pair source. In addition, we also observed antibunching behavior
between the fundamental mode photons and subharmonic mode photons.


\begin{thebibliography}{99}
\bibitem{TPM} W. Denk, J. H. Strickler, and W. W. Webb, Science \textbf{248}%
, 73--76 (1990).

\bibitem{MPA} J. W. Hudgens, T. G. DiGiuseppe, and M. C. Lin, Chem. Phys.
\textbf{79}, 571 (1983).

\bibitem{QMtest} A. Aspect, P. Grangier, and G. Roger, Phys. Rev. Lett.
\textbf{47}, 460 (1981).

\bibitem{QCM} D. Bouwmeester, J. W. Pan, K. Mattle, M. Eibl, H. Weinfurter,
and A. Zeilinger, Nature \textbf{390}, 575 (1997).

\bibitem{Qcomputer} P. Kok, W. J. Munro, K. Nemoto, T. C. Ralph, J. P.
Dowling, G. J. Milburn, Rev. Mod. Rhys. \textbf{79}, 135 (2007).

\bibitem{16} C. Xiong, C. Monat, A. S. Clark, C. Grillet, G. D. Marshall, M.
J. Steel, J. Li, L. O'Faolain, T. F. Krauss, J. G. Rarity, and B. J.
Eggleton, Opt. Lett. \textbf{36}, 3413 (2011).

\bibitem{SPDC} J. W. Pan, Z. B. Chen, C. Y. Lu, H. Weinfurter, A. Zeilinger,
and M. \.{Z}ukowski, Rev. Mod. Phys. \textbf{84}, 777 (2012).

\bibitem{OPO} U. Herzog, M. Scholz, and O. Benson, Phys. Rev. A \textbf{77}, 023826
(2008).

\bibitem{fiber} O. Cohen, J. S. Lundeen, B. J. Smith, G. Puentes, P. J.
Mosley, and I. A. Walmsley, Phys. Rev. Lett. \textbf{102}, 123603 (2009).

\bibitem{atom} B. Srivathsan, G. K. Gulati, B. Chng, G. Maslennikov, D.
Matsukevich, and C. Kurtsiefer, Phys. Rev. Lett. \textbf{111}, 123602 (2013).

\bibitem{Qnetwork} H. J. Kimble, Nature \textbf{453}, 1023 (2008).

\bibitem{PB} A. Imamo\={g}lu, H. Schmidt, G. Woods, and M. Deutsch, Phys.
Rev. Lett. \textbf{79}, 1467 (1997).

\bibitem{UPB} M. Bamba, A. Imamo\u{g}lu, I. Carusotto, and C. Ciuti, Phys.
Rev. A \textbf{83}, 021802 (2011).

\bibitem{UPB1} H. Flayac and V. Savona. Phys. Rev. A \textbf{96}, 053810 (2017).

\bibitem{TLS_PB} Xinyun Liang, Zhenglu Duan, Qin Guo, Cunjin Liu, Shengguo Guan, and
Yi Ren, Phys. Rev. A \textbf{100}, 063834 (2019).

\bibitem{TLS2} Xinyun Liang, Zhenglu Duan, Qin Guo, Shengguo Guan, Min Xie, and C. J. Liu, Phys. Rev. A \textbf{102}, 053713 (2020)

\bibitem{EIT} M. Fleischhauer, A. Imamoglu, and J. P. Marangos, Rev. Mod.
Phys. \textbf{77}, 633 (2005)

\bibitem{CPBEx} K. M. Birnbaum, A. Boca, R. Miller, A. D. Boozer, T. E.
Northup, and H. J. Kimble, Nature \textbf{436}, 87--90 (2005).

\bibitem{QDC} J. Kasprzak, S. Reitzenstein, E. A. Muljarov, C. Kistner, C.
Schneider, M. Strauss, S. H\"{o}fling, A. Forchel, and W. Langbein, Nat.
Mater. \textbf{9}, 304--308 (2010).

\bibitem{SQS} J. M. Fink, M. G\"{o}ppl, M. Baur, R. Bianchetti, P. J. Leek,
A. Blais and A. Wallraff, Nature \textbf{454}, 315--318 (2008).

\bibitem{OMS} D. Y. Wang, C. H. Bai, S. Liu, S. Zhang, and H. F. Wang, New
J. Phys. \textbf{22}, 093006 (2020).

\bibitem{OMS1} D. Y. Wang, C. H. Bai, X. Han, S. Liu, S. Zhang, and H. F.
Wang, Opt. Lett. \textbf{45}, 2604 (2020).

\bibitem{SONSM} W. T. M. Irvine, K. Hennessy, and D. Bouwmeester, Phys. Rev.
Lett. \textbf{96}, 057405 (2006).

\bibitem{SONSM1} A. Majumdar and D. Gerace. Phys. Rev. B \textbf{87}, 235319 (2013).

\bibitem{SONSM2} D. Gerace and V. Savona. Phys. Rev. A \textbf{89}, 031803 (2014).

\bibitem{SONSM3} M. Li, Y. L. Zhang, H. X. Tang, C. H. Dong, G. C. Guo, and
C. L. Zou, Phys. Rev. Appl. \textbf{13}, 044013 (2020).

\bibitem{SPDC0} Y. H. Zhou, X. Y. Zhang, Q. C. Wu, B. L. Ye, Z. Q. Zhang, D.
D. Zou, H. Z. Shen, and C. P. Yang, Phys. Rev. A \textbf{102}, 033713 (2020).
\end{thebibliography}
\end{document}